# Physical mechanisms of timing jitter in photon detection by current carrying superconducting nanowires

Mariia Sidorova, Alexej Semenov, Heinz-Wilhelm Hübers
*DLR Institute of optical systems, Rutherfordstrasse 2, 12489 Berlin, Germany*

Ilya Charaev, Artem Kuzmin, Steffen Doerner, Michael Siegel
*Institut für Mikro- und Nanoelektronische Systeme (IMS), Karlsruher Institut für Technologie (KIT), Hertzstr. 16, 76187 Karlsruhe, Germany*

We studied timing jitter in the appearance of photon counts in meandering nanowires with different fractional amount of bends. Timing jitter, which is the probability density of the random time delay between photon absorption in current-carrying superconducting nanowire and appearance of the normal domain, reveals two different underlying physical scenarios. In the deterministic regime, which is realized at large currents and photon energies, jitter is controlled by position dependent detection threshold in straight parts of meanders and decreases with the current. At small photon energies, jitter increases and its current dependence disappears. In this probabilistic regime jitter is controlled by Poisson process in that magnetic vortices jump randomly across the wire in areas adjacent to the bends.

## I. INTRODUCTION

In the interaction between photons and matter, conversion of photon energy to excitation in the electronic spectrum and further to measurable change of any of macroscopic parameters is the subject of statistical fluctuations. These fluctuations randomize the time delay between photon absorption and the appearance of the change in the parameter of interest. In each particular case, revealing physical mechanisms which constitute the delay improves understanding of light-matter interaction. Besides that, random time delay causes timing jitter in the appearance of the voltage transient which a photon detector produces to signal detection event. The measure of the timing jitter is the width of the statistical distribution in the arrival times of voltage transients with respect to the corresponding photon absorption times. Nowadays, one of the actively developing detector technologies is superconducting nanowire single-photon detectors (SNSPDs). Although impressive progress has been achieved during last decades in SNSPD technology, performance of these detectors is still improving. Timing jitter is one of the SNSPD metrics which is important for many applications and may have great potential for extending application field of SNSPDs. The latter is true because relative contributions of different physical mechanisms to the magnitude of the timing jitter are not fully understood. In SNSPD applications, timing jitter limits the accuracy of measurements of photon arrival times. Hence, it plays a crucial role in laser ranging, communication technologies or time-resolved correlation measurements.

During last decade many groups have reported on small timing jitter in SNSPDs. However, measurements were done with different detector layouts and electronics that hamper direct comparison. System jitter with full width at half maximum (FWHM) as low as 18 ps has been demonstrated for SNSPD based on NbN [1, 2]. It has been realized that electronic noise severely enhances system jitter and introduces the current dependence of the jitter. Subtracting the noise contribution one obtains jitter inherent to the detector itself which is called intrinsic jitter. It was found that the intrinsic jitter increases in nanowires with smaller thickness and larger kinetic inductance per unit length [3]. Furthermore, the jitter increases with the size of the detector [4, 5] and is less for the central part of the detector area as compared to peripherals [6]. Although low jitter itself is a challenge, it becomes attractive only in conjunction with the practical values of the detection efficiency and maximum count rate. Since the size of the detector affects differently these two metrics, the size stays necessarily in the list of trade-off parameters. Jitter increases in nanowires from superconducting films with low transition temperature. For nanowires from WSi [7] and MoSi [8], jitter is almost one order of magnitude larger than in nanowires from NbN.

While instrumental aspects of the system timing jitter have been thoroughly discussed, physical mechanisms of the intrinsic jitter remain largely unclear. Revealing those mechanisms should give the answer how to decrease jitter value and what is the limit. For WSi, Fano fluctuations were shown to broaden the decay of the photon detection efficiency with the decrease of the current trough the nanowire [9]. They should also affect the time delay between photon absorption and the emergence of the resistive state. Whether this mechanism may affect jitter in NbN nanowires is not clear. The delay of the resistive stay in NbN was found to exceed 65 ps [10]. It has been shown that this delay may depend on the film thickness via the escape of non-thermal phonons to the substrate [11]. Recently, spread in the traveling time of a magnetic vortex across the nanowire was analyzed [12]. For NbN, it can contribute a few picoseconds to the intrinsic timing jitter.

TABLE 1. Summary of different contributions to the intrinsic jitter, affecting factors and favoring experimental techniques.

| Jitter source | Factors affecting jitter | | | Quantitative description | Experimental approach |
|---|---|---|---|---|---|
| | External | Internal | | | |
| Geometric | | Kinetic inductance, electrical path | | Transmission line | Differential technique |
| Local | Operational conditions:<br>• temperature,<br>• bias current,<br>• photon energy | Location of the photon absorption:<br>• Straight wire<br>• Bend<br><br>Detection regime:<br>• Single-photon<br>• Multi-photon | Detection scenario: | • Deterministic | Diffusive normal-core | Standard technique |
| | | | | • Probabilistic | Vortex crossing | |

In terms of the probability theory, arrival of the voltage transient at the recording instrument, i.e. photon counting event, is a composite sequential event characterized by the delay time after photon absorption. The delay time is the continuous random variable with the probability density function (PDF) presented by the histogram of arrival times. In a superconducting nanowire, photon counting event is a sequence of several elementary events (stages), which are (i) photon absorption, (ii) thermalization i.e. energy transfer from the absorbed photon to electrons in the nanowire, (iii) appearance of the hot-spot, (iv) emergence of the normal domain, and (v) propagation of two current steps through the nanowire and ground plane, respectively, to the input of the common transmission line. All these stages contribute to the intrinsic jitter which is associated with the nanowire itself. Table I summarizes different factors affecting components of intrinsic jitter. Any experimentally measured histograms of the arrival times bear additionally contributions from the experimental environment [13] via noise in electronics, dispersion in optics and random time difference between laser pulse and reference signal.

Local jitter inheres in the photon detection itself. It accumulates contributions from first four stages listed above. With the proper polarization, absorbance is uniform over the nanowire and does not contribute extra randomness to the delay time. Thermalization (ii) is the subject of Fano fluctuations [14]. They randomize the amount of energy transferred from the photon to electrons and, consequently, the size of the hot-spot when it is defined at the fixed delay time or the delay time in the appearance of the hot-spot with the fixed size. Given that the rate of dark counts is sufficiently small, the hot-spot (iii) opens the time window for vortex crossing. Following the location of the absorption site, vortex crossing occur either in the straight portion of the nanowire or in the bend. These are two mutually exclusive events for which total PDF is a sum of individual PDFs [15] weighted with relative areas covered by bends and straight parts.

It is commonly accepted [16, 17] that the normal zone emerges (iv) when a magnetic vortex crosses the nanowire. Vortex crossing obey either deterministic or probabilistic scenario. They both randomize the total delay time. Their individual PDFs include contributions from the random flight-time of the vortex across the nanowire [12] and from the random start time of the crossing. While the former is the same for both scenarios, the latter is controlled by the instant value of the potential barrier for vortex crossing and hence depends on the photon energy. Opening of the time window by the hot-spot and vortex crossing are independent sequential events. PDF of the composite event is the integral of the product of elementary PDFs with the fixed time difference [15]. Which detection regime is realized at particular absorption site depends on the operation parameters: temperature, current, and photon energy. In both detection scenarios, additional jitter arises from the position dependence of the detection probability [18] and, correspondingly, detection current. The local detection current is the smallest current through the nanowire that is required to achieve 100% detection probability for photon absorbed at particular location [19, 20].

Vortex crossing of any kind generates the normal domain and, consequently, two current steps which propagate to opposite directions from the absorption site. Generally, the propagation time is different for different absorption sites. This introduces geometric contribution to the PDF of the total delay time. In the framework of the transmission-line approach, geometric jitter is controlled by the length of nanowire, its kinetic inductance, and the layout which define jointly propagation velocity of electrical signal along the nanowire (Table 1). Experimental approach which allows for direct measurements of the geometric jitter is called differential technique [4]. In this technique, arrival times of two electrical pulses originating from the same count event are measured independently at two ends of a nanowire.

Here we study intrinsic jitter in differently shaped NbN meanders at two wavelengths 800 and 1560 nm which cover both deterministic and probabilistic detection scenarios. Our meanders contain different relative amount of straight wires and fixed number of bends. Along with the transmission line approach, a set of different layouts allows us to evaluate local and geometric contributions to the jitter. Furthermore, we estimate separately contributions from straight nanowires and bends. We show that jitter statistics changes with the wavelength and is also different for straight nanowires and for bends. We discuss mechanisms which cause nonmonotonic jitter behavior with the current in the deterministic detection regime and monotonic jitter dependence on the current in the probabilistic detection regime. We propose simplified physical models which provide good quantitative description of our experimental findings.

## II. EXPERIMENT

### A. Sample preparation

Our meanders were drawn by electron beam lithography from 5 nm thick NbN superconducting films deposited on $Al_2O_3$ substrate. Nanowires had a width of 90 nm and a filling factor of 50% (Fig.1). The meandering nanowire was connected to contact pads shortening a coplanar transmission line. We studied meanders having different sizes: 4 x 4, 4 x 3, 4 x 2, and 4 x 1 $\mu m^2$ but the same number of bends. Their shape and size were identical for all meanders as well as within one meander. Correspondingly, the length of straight wires (*L*) between bends varied as 3.5, 2.5, 1.5, 0.5 $\mu m$. Hence, in the biggest meander (Fig. 1), the nanowire consists mostly from straight parts while in the smallest meander (inset to Fig. 1) bends dominate. Transport measurements at 4.2 K showed similarity in the values of critical currents 37.6 – 41.1 $\mu A$ and superconducting transition temperatures 13.05 – 13.35 K of different meanders. Normal square resistance of the original films was $R_S$ = 243 Ohm/sq at 25 K.

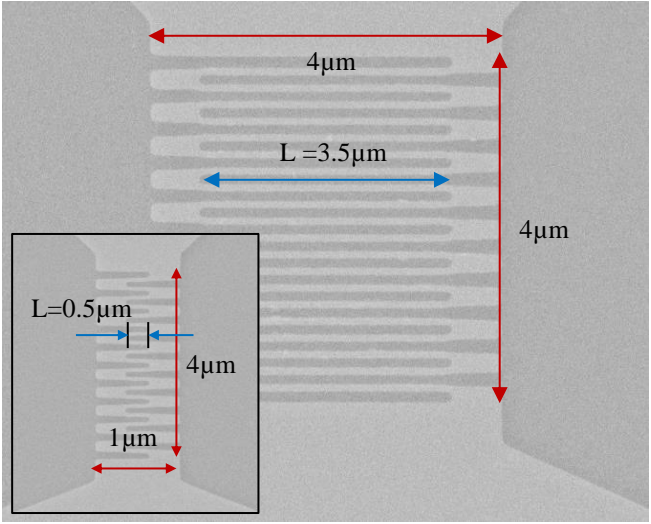

FIG. 1. (Color online) SEM image of the largest meander. The size of the meander is 4x4 $\mu m^2$ and the length of straight wires between bends is 3.5 $\mu m$. Inset shows the smallest meander, the size of which is 4x1 $\mu m^2$ and the length of straight wires between bends is 0.5 $\mu m$.

### B. Experimental approach

Jitter measurements were carried out at 4.2 K and two wavelengths 800 nm and 1560 nm. Meanders were uniformly illuminated by laser pulses with sub-picosecond duration with a repetition rate between 80 MHz and 100 MHz. Meanders were biased via a bias-tee with a DC current supplied by a battery-powered electronics. Voltage transients generated by counting events were amplified with a room temperature low noise amplifier, which had the bandwidth from 100 MHz to 8 GHz and the noise level of 1.4 dB, and acquired with 50 GHz sampling oscilloscope. The scope was triggered by electrical pulses from a fast photodiode which was illuminated by laser pulses (Fig. 2(a)). A typical voltage transient obtained by sampling many counting events is shown in Fig. 2(b). To build a histogram, we accumulated more than $10^4$ points inside an acquisition window on the rising edge of the voltage transient (shown by the rectangle in Fig. 2(b)). We associated the arrival time of the transient with the time when the points from this transient appear within the window. The distribution of arrival times (histogram) was then computed with the time bin less than 0.3 ps. The histogram represents probability density function of the arrival time which is considered to be a random continuous variable. Extracted PDFs (one of them exemplarily shown in the inset in Fig. 2(b)) typically have a non-Gaussian profile with a tail extended to larger arrival times. Therefore, we measured the full width at half maximum (FWHM) of a histogram and defined the standard deviation $\sigma_{syst}$ as the 1/2.35 part of the measured value. Hereafter we will use the standard deviation as the measure of timing jitter. We have found that the system jitter, $\sigma_{syst}$, was strongly affected by the noise in the electrical network, fluctuations in the transient amplitude, and the level, *H*, (Fig. 2(b)) where the acquisition window was positioned. To minimize acquisition time and avoid slow drifts in the ambient conditions, we set the height of this window at the largest value that had not yet affected the extracted jitter.

We estimate instrumental contribution to the measured jitter substituting the meander with another fast photodiode and measuring histograms with different optical fibers between the laser and the second photodiode. All instrumental histograms had Gaussian form and were fully symmetric down to the level of $10^{-3}$ from the distribution maximum. The instrumental jitter $\sigma_{instr}$ was less than 1.5 ps for illumination via open beam and increased to 2 ps or 1.7 ps when light was delivered to the photodiode by two meters of multimode fiber or three meters of single mode fiber, respectively. Noise contribution to the system jitter was estimated [3, 13] as $\sigma_{noise} = \sigma U_N \cdot \tau / A_{mean}$ where $A_{mean}$ is the mean transient amplitude, $\tau$ is duration of the rising edge of the transient and $\sigma U_N$ is standard deviation extracted from the height of histogram of sampling points at the base line (Fig. 2 (b)).

It is commonly known that SNSPDs pulses exhibit amplitude fluctuations. Fig. 3(a) schematically shows two voltage pulses with different amplitudes arrived at the same time. If one builds up time distribution of sampling points at the level $H$, any difference in the amplitudes will broaden this distribution producing artificial jitter, $\sigma_{amp}$. Simple math results in the following connection between $\sigma_{amp}$ and the standard deviation $\sigma U_A$ in the distribution of transient amplitudes

$$\sigma_{amp} = \sigma U_A \frac{\tau H}{A_{mean}^2}, \quad (1)$$

where $\sigma U_A = \sqrt{\sigma U_\Sigma^2 - \sigma U_N^2}$, and $\sigma U_\Sigma$ is the standard deviation in the vertical distribution of point heights measured as the top of the voltage transient (Fig. 2 (b)).

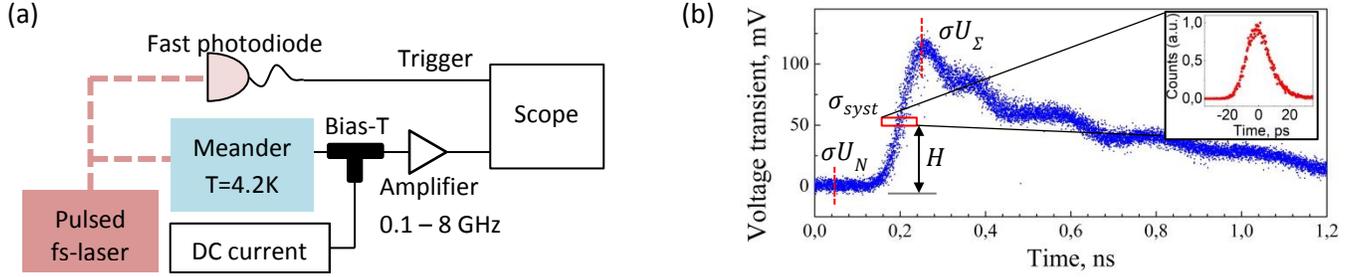

FIG. 2. (Color online) (a) Schematics of the setup for jitter measurements at 800 nm and 1560 nm. (b) Voltage transient recorded by the 50 GHz sampling oscilloscope. Points occurring within the rectangle window at the level $H$ are used to build statistical distribution of arrival times shown in the inset. Dotted lines show locations where the vertical distributions of sampling points were additionally measured. $\sigma U_\Sigma$, $\sigma U_N$ and $\sigma_{syst}$ denote corresponding standard deviations.

We found that electrical noise and amplitude fluctuations had almost Gaussian distributions. Further assuming that they are statistically independent, we obtained the standard deviation in the intrinsic jitter as

$$\sigma_{int} = \sqrt{\sigma_{syst}^2 - (\sigma_{noise}^2 + \sigma_{amp}^2 + \sigma_{instr}^2)}. \quad (2)$$

Appearance of the normal domain and the propagation of the current steps are also sequential independent events. Although the shape of experimental PDFs (experimental histograms) differ from the normal distributions we suppose that geometric PDF and PDF of the normal domain are statistically stable and the dispersion in the intrinsic PDF is the sum of dispersions in these partial PDFs. Since statistics in the appearance of the normal domain is inherent to the absorption site, we will hereafter denote the corresponding local standard deviation as $\sigma_{loc}$ and the geometric contribution as $\sigma_{geom}$. The standard deviation in the intrinsic jitter can be then presented as

$$\sigma_{int} = \sqrt{\sigma_{loc}^2 + \sigma_{geom}^2}. \quad (3)$$

We have to note here that our experimental approach effectively eliminates geometric jitter. One end of the nanowire of the total length, $L$, is connected to the ground plane of the coplanar line as it is schematically shown in Fig. 3(b). When a photon initiates counting event at the distance $x$ from this end of the nanowire, two current steps propagate to the opposite directions from the absorption site. One arrives at the common coaxial input after the time $(L-x)/v$ where $v$ is the propagation velocity of the current step in the nanowire. Another step propagates till the shorted end of the nanowire, reflects in the ground plane and further travels to the common coaxial input via the ground plane. This second step arrives at the common input the time $x/v + L/v^*$ where $v^*$ is the propagation velocity of the current step in the ground plane. The difference between arrival times is less than the time resolution of the common amplifier. Therefore it sees the sum of two steps. The arrival time of the sum is associated with the arithmetical mean of the two times $L(v + v^*)/(2\, v\, v^*)$ which does not depend on the position of the absorption site. Geometric jitter may appear in this configuration only if the damping and dispersion in the nanowire and in the ground plane are different. Hence, we expect small geometric jitter if any to be present in our experimental data.

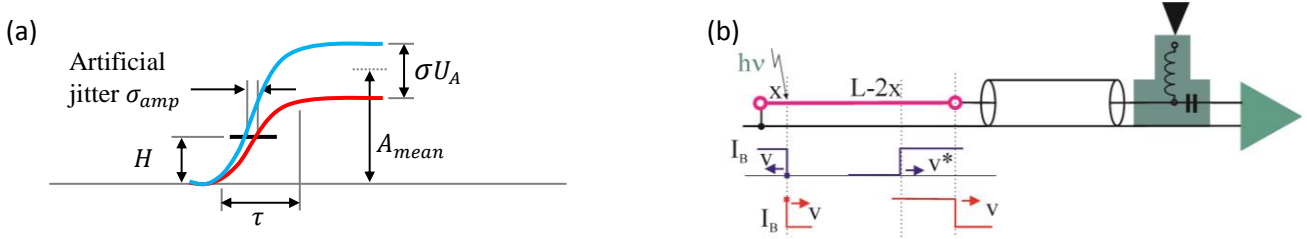

FIG. 3. (Color online) (a) Artificial jitter $j_{amp}$ due to the difference in amplitudes of two pulses arriving at the same time. (b) Schematics of the appearance of the geometric jitter. Nanowire with the total length, $L$, is to the common coaxial input and to the ground plane. Photon absorbed at the distance $x$ from the end of the nanowire initiates two current steps propagate in the opposite directions from the absorption site. Although they arrive at the common input at different times, the mean arrival time does not depend on the position of the absorption site.

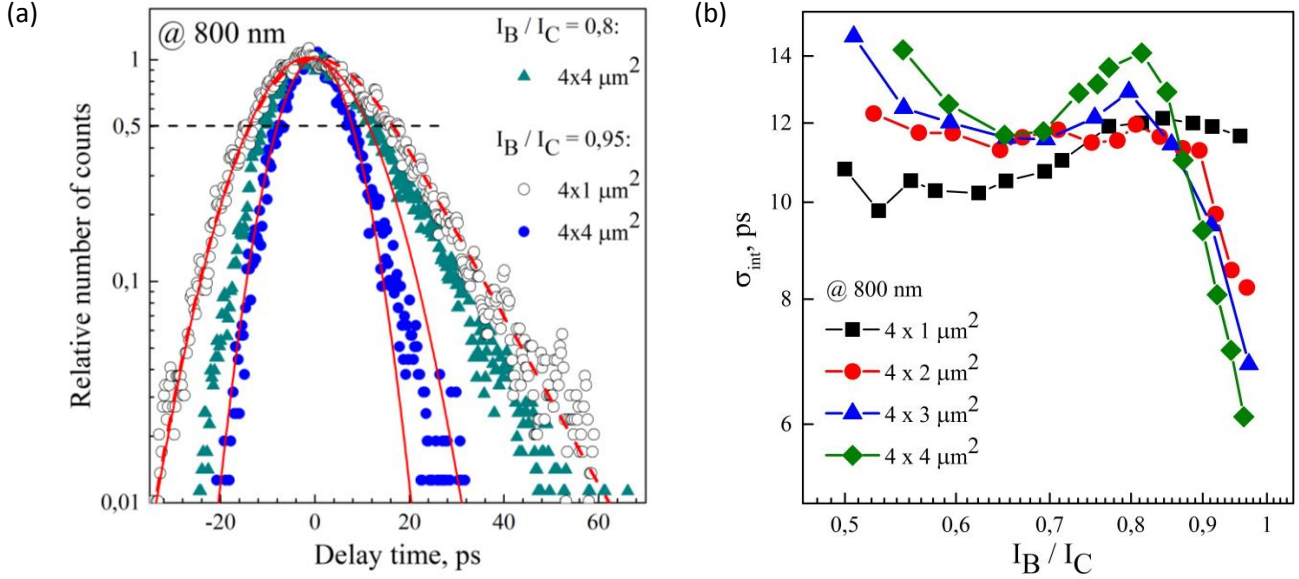

FIG. 4. (Color online) (a) Histograms (PDFs) of the delay time for the meander 4 x 4 μm² (closed symbols) at currents 0.95 $I_C$ (circles) and 0.8 $I_C$ (triangles) and for the meander 4 x 1 μm² at the current 0.95 $I_C$ (open symbols). Data were obtained at 4.2 K and the wavelength 800 nm. Horizontal dashed line marks the level where FWHM were measured. Solid lines show Gaussian fits to the left parts of histograms. Dashed line shows the best fit of the PDF at 0.8 $I_C$ with Eq. (4). (b) Standard deviation in the intrinsic jitter as function of bias current for four studied meanders. Lines are to guide the eyes.

## III. RESULTS

### A. Excitation wavelength 800 nm

Experimental histograms, i.e. relative numbers of photon counts per time-bin are shown in Fig. 4(a) for two meanders as function of the delay time. The data were acquired at the wavelength 800 nm and at the bias current $I_B = 0.95\ I_C$ where $I_C$ is the experimental critical current. As discussed above, the data represent probability distribution functions of arrival times. Since the exact delay time between the photon arrival and the transient appearance is not known, the maxima of PDFs were assigned zero delay values. We are not going to discuss here the true probability of the photon detection. Therefore, PDFs are left non-normalized. In the semi-logarithmic scale, asymmetry in PDFs is clearly seen. Such deviation from normal distribution is typical for meanders and was observed by a number of groups [1, 2, 6, 13]. The asymmetry is more pronounced for the meander with smaller relative amount of straight wires. The cumulative distribution has Gaussian shape at small delay and drops linearly (in semi-logarithmic scale) with time at large delays. Decrease in the bias current much stronger affects the dispersion at large delays than at small delays, although the shapes remain unchanged. Linear decrease of the PDFs at large delays evidences that the cumulative delay includes a stage with exponential distribution of probability density. The asymmetry rules out Gaussian fit (solid lines in Fig. 4(a)) as a valid instrument of finding standard deviation in PDFs. Instead, we measured the full width at half maximum (FWHM) of each experimental PDF and associated the standard deviation with 1/2.35 part of the measured value. We further subtracted instrumental, fluctuation and noise contributions according to equations (1 – 3) to obtain intrinsic jitter. We have found that $\sigma_{amp}$ and $\sigma_{noise}$ significantly affect $\sigma_{syst}$ only at small bias currents. For instance, for the meander 4 x 1 μm² at 0.95 $I_C$ the value of $(\sigma_{noise}^2 + \sigma_{amp}^2)^{1/2} \approx 2.6$ ps against $\sigma_{syst} \approx 12.2$ ps. However, at the bias current 0.6 $I_C$ these values are ≈ 8.7 ps and ≈ 12.7 ps, respectively. Instrumental contribution to $\sigma_{syst}$ was negligibly small and current independent. Since the instrumental contribution and the noise contribution were symmetric and had Gaussian profiles, we suppose that the observed asymmetry is inherent to photon detection in meanders.

The intrinsic jitter is shown in Fig. 4(b) as function of the bias current for four studied meanders. At small currents, the jitter increases with an increase in the total length of straight wires. Contrarily, the intrinsic jitter turns to have the smallest value for the largest meander when the bias current approaches the critical current. By measuring the dependences of the count rate on the light intensity and the bias current, we verified that at currents less than 0.7 $I_C$ meanders undergo the transition to multi-photon detection regime. Jitter for multiphoton detection goes beyond the scope of the present study. Here we will discuss results exclusively at large bias currents.

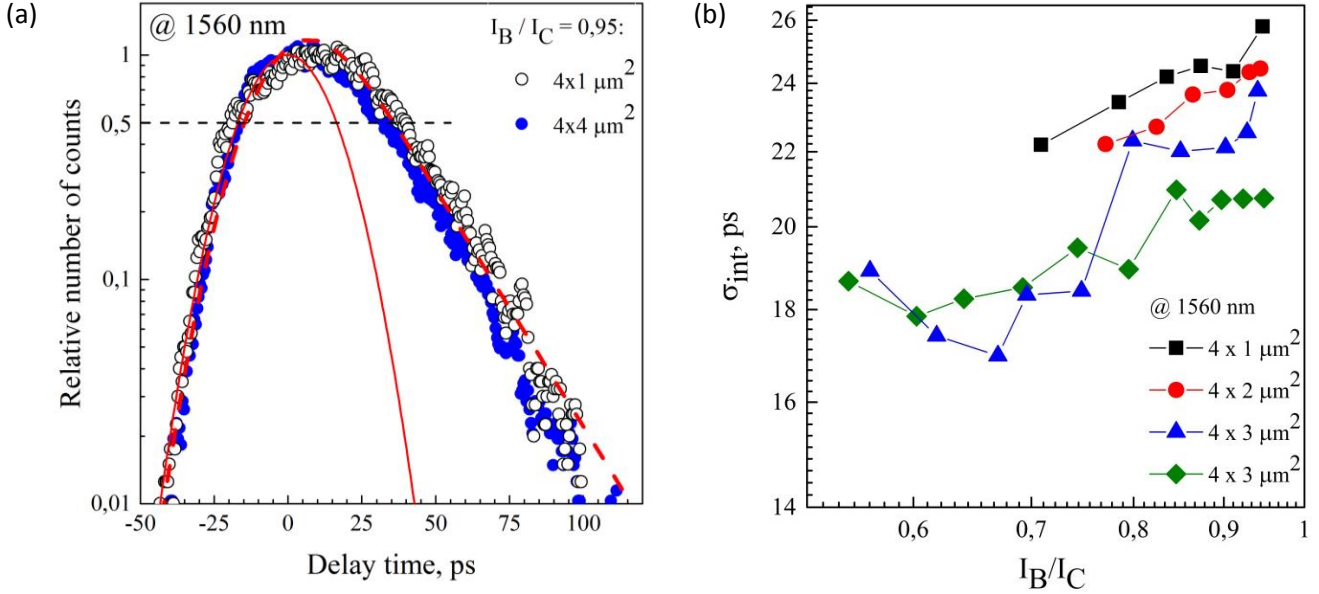

FIG. 5. (Color online) (a) Histograms (PDFs) of the delay time at the current 0.95 $I_C$ for the meanders 4 x 4 µm² (closed symbols) and for the meander 4 x 1 µm² (open symbols). Data were obtained at 4.2 K and the wavelength 1560 nm. Horizontal dashed line marks the level where FWHM were measured. Solid line shows Gaussian fit to the left parts of the histograms. Dashed line shows the best fit of the PDF with Eq. (4) for the meander 4 x 1 µm². (b) Standard deviation in the intrinsic jitter as function of bias current for four studied meanders. Lines are to guide the eyes.

### B. Excitation wavelength 1560 nm

Histograms acquired at 1560 nm for two different meanders at $I_B = 0.95\, I_C$ are shown in Fig. 5(a). At this wavelength histograms were also asymmetric but the degree of asymmetry was larger than in histograms acquired at 800 nm. Differently to the wavelength 800 nm, all meanders demonstrated the same dispersion at small delays. At large delays, dispersion increases with the decrease of the meander size. For all meanders, intrinsic jitter monotonically increases with the bias current as it is shown in Fig. 5(b). Due to low detection efficiency at 1560 nm, specifically for smaller meander, the required integration time becomes larger than long-term stability time of the setup. As the result, the instantaneous dispersion first decreases with the increase in the integration time but starts to increase when the integration time exceeds the long-term stability time. Therefore, we were not able to get reliable data for small meanders at small bias currents. Anyway, data at currents less than 0.7 $I_C$ would not be discussed here since multiphoton detection dominates at small currents.

### IV. DISCUSSION

#### A. Probability density functions: Histograms

As it has been explained earlier the photon count is a composite event including several sequential stages. We first estimate their expected contributions to the intrinsic jitter. According to the original study [14], for statistically independent scattering events the variance (standard deviation) in the quantum yield, $\sigma N$, depends on the total deposited energy, $E$, and the mean energy per particle, $\varepsilon$, as $\sigma N = (FE/\varepsilon)^{1/2}$, where $F = 0.2 - 0.3$ is the Fano factor. For electron avalanche in a superconductor, $\varepsilon$ equals the superconducting energy gap, $\Delta$, while $E$ is the photon energy, $h\nu$, reduced by the additional factor $a < 1$ which stays for the mean effectiveness of the energy transfer from the photon to electrons. Fluctuations in the quantum yield set the ultimate value of the energy resolution in tunnel-junction photon detectors. Despite extensive efforts this limit has never been achieved [21]. Due to fluctuations of different origins, experimental energy resolution was typically one order of magnitude worse. In our case, the size of the two-dimensional diffusive hot spot is proportional to its growth time, $\tau$, and to the logarithm of the total number of non thermal electrons, $N$. Hence the variance in the growth time $\sigma\tau = \tau\, \sigma N/N = (\alpha F\Delta/(h\nu))^{1/2}$. With typical parameters of NbN nanowires, $\sigma\tau/\tau \approx 10^{-3}$ and the variance in the growth time drops below one picosecond even for the largest reported $\tau \approx 20\, ps$ [22]. We therefore neglect the contribution of Fano fluctuations in the following consideration.

We next suppose that our experimental configuration eliminates completely geometric contribution to the intrinsic jitter. The magnitude of the geometric contribution is indeed negligible as it will be shown below. The photon count is then a combination of two sequential and statistically independent events: (1) opening of the time window for vortex crossing by the hot-spot and (2) vortex crossing itself. In terms of probability theory such composite event is described by a probability density function $h(t) = \int f_1(\tau) f_2(t - \tau) d\tau$ where $f_1(t)$ and $f_2(t)$ are PDFs of these two sequential events [15].

We assume that the opening time obeys normal distribution with the mean value $\mu$, which we set to zero for simplicity, and the variance $\sigma$, which approximately equals the lifetime of the hotspot. Possible reasons for the distribution of the opening time will be discussed below. One of them is the position dependence of the detection current [12, 18-20]. The value of the hotspot lifetime depends on how the borders of the spot are defined and may vary between electron thermalization time ≈7 ps [23] and the diffusion time across the nanowire. The latter is approximately 20 ps for the 100 nm wire from NbN. Vortex crossing is a Poisson process in that the time delay between the opening of the window and the start time of a first successful crossing obeys the same statistics as e.g. nuclear decay or short noise. The PDF of the appearance time of the first event is an exponential function with the characteristic time. The reciprocal of this time represents simultaneously the mean crossing rate and its variance. The flight time of the vortex across the nanowire is also statistical variable with the mean value $\tau_v = \Phi_0(w/\xi)^2(2\pi I_B R_S)^{-1}$ where $w$ and $\xi$ are the width of the wire and the coherence length, respectively, and $\Phi_0$ is the magnetic flux quantum. With typical NbN parameters the flight time amounts at 12 ps at the experimental critical current. We will not distinguish at this stage between the delay of the crossing and the time of crossing and approximate both by the single exponential PDF with the mean value $\tau_0$. The PDF of the composite event is the one known as exponentially modified Gaussian distribution

$$h(t) = \frac{1}{2\tau_0}\exp\left(\frac{1}{2\tau_0}(\frac{\sigma^2}{\tau_0} - 2t)\right) \cdot \left(1 - \mathrm{erf}\left(\frac{(\frac{\sigma^2}{\tau_0} - t)}{(\sigma\sqrt{2})}\right)\right), \quad (4)$$

where $\mathrm{erf}(x)$ is the error function. We used the Eq. (4) to fit our experimental histograms. Result for the 4 x 4 μm² meander at wavelengths 800 nm and 1560 mm are shown in Figs. 4(a) and 5(a), respectively. Table 2 summarizes the best fit parameters for different meanders and wavelengths.

Comparing fit parameters from the Table 2 with the experimentally measured values of the intrinsic jitter $\sigma_{int}$ (Figs. 4(b) and 5(b)), we found that for both wavelengths they satisfy the relation $\sigma_{int} = (\sigma^2 + \tau_0^2)^{1/2}$ with the accuracy better than 10% in the whole range of currents. This observation supports the validity of the simplified procedure which we used to extract the magnitude of the intrinsic jitter from experimental histograms. The observation also confirms theoretical prediction that the dispersion of a composite event combining several sequential and statistically independent events represents the sum of dispersions of the elementary events.

Data in Table 2 show that exponential PDF ($\tau_0$) dominates intrinsic jitter at large wavelengths and small currents. Available experimental observations evidence, that photon detection at this range of parameters is better described by probabilistic detection scenario. The spectral cut-off in the detection efficiency of the 4 x 4 μm² meander occurs around the wavelength 800 nm [24]. Detection at smaller wavelengths was associated with the deterministic scenario in that the bias current is larger than the local value of the detection current at any location in the meander [25]. Well beyond the cut-off, when the current is smaller than the smallest detection current in the meander, detection obeys probabilistic scenario. Photon detection beyond the threshold was associated with the dark count event localized at the absorption site [16]. The absorbed photon reduces locally the barrier for vortex crossing to the extent that crossing at the absorption site has larger probability to happen than anywhere else in the nanowire. Blurred transition between two regimes corresponds to separation of the meander into two parts undergoing different detection scenarios. Hence, we suppose that the exponential elementary PDF corresponds to the probabilistic vortex crossing. It is also seen from Table 2 that the probabilistic fraction of the intrinsic jitter increases with the increase of the relative length of bends in the total length of meanders. In the next section we show that bends are primarily responsible for exponential contribution to the intrinsic jitter.

TABLE 2. Best fit parameters for fits of experimental histograms (probability density functions) with Eq. (4).

| | $\frac{I_B}{I_C}$ | Fitting parameter (ps) | Meander size (μm²) | | | |
|---|---|---|---|---|---|---|
| | | | 4 x 4 | 4 x 3 | 4 x 2 | 4 x 1 |
| @ 800 nm | 0.95 | $\sigma$ | 6.1 | 6.2 | 7 | 10 |
| | | $\tau_0$ | 4 | 5.3 | 8.3 | 11.5 |
| | 0.8 | $\sigma$ | 7.1 | 8.2 | 8.4 | 9 |
| | | $\tau_0$ | 10 | 11 | 11.3 | 12.5 |
| @ 1560 nm | 0.95 | $\sigma$ | 14 | 14 | 14 | 14 |
| | | $\tau_0$ | 18 | 19.5 | 20 | 20.5 |

### B. Local jitter: Bends and straight wires

In an attempt to separate counts arriving from bends and straight wires, we geometrically divide each meander in two parts. The part with bends has the length $L_b = n(2w + s)$ where $s$ is the separation between wires and $n$ the number of bends. The straight wires have total length $L-L_b$ where $L$ is the total length of the meandering wire. Photon counts coming from bends and wires are mutually exclusive independent events and the probability of either to occur is the sum of the probabilities of their occurrences taken with corresponding geometric weights $F = L_b/L$ and $1-F$. The dispersion for two mutually exclusive events is the sum of dispersions of elementary events. According to Eq. (1), intrinsic timing jitter contains statistically independent contributions from local and geometric jitters. We further present local jitter as containing statistically independent local jitter from bends, $\sigma_{bend}$, and from straight wires, $\sigma_{wire}$. Considering the nanowire as a portion of a transmission line, we present geometric jitter as $L/k$, where $k$ plays the role of the effective propagation velocity of electric transients. Eq. (1) can be now rewritten to obtain

$$\sigma_{int} = \sqrt{F\,\sigma_{bend}^2 + (1-F)\,\sigma_{wire}^2 + (L/k)^2}. \quad (5)$$

Writing down this equation for each detector, we obtain a system of four equations with three unknown independent variables $\sigma_{bend}$, $\sigma_{wire}$ and $k$. We found solutions for four possible combinations each containing three equations from the set of four. The solution for the wavelength 800 nm and $I_B = 0.95\,I_C$ delivers $k = 70$ μm/ps and the geometric jitter varying between 0.3 ps and 1.17 ps for meanders from 4 x 1 μm² to 4 x 4 μm², correspondingly. The value of the effective velocity is only one half of the expected propagation velocity ($\approx 140$ μm/ps) for a coplanar line on sapphire. This result confirms our supposition that the geometric jitter is effectively eliminated by our experimental technique. Since the geometric contribution is much smaller than any measured value of the intrinsic jitter, we neglected geometric jitter in the following consideration. Solving systems with all possible pairs of 4 equations for different currents and wavelengths, we obtain $\sigma_{wire}$ and $\sigma_{bends}$ with the accuracy better than 12% (Fig. 6). Parameter F was assumed to be current independent. We verified that changing this parameter leads to changes in absolute values of $\sigma_{wire}$ and $\sigma_{bends}$ but does not affect their current dependences. Fig. 6 shows current dependences of the events originating in bends and wires. Jitter from bends grows monotonously with the current at both wavelengths. Jitter from wires behaves differently. It demonstrates monotonous growth with the current only at the larger wavelength. At smaller wavelengths, sharp decrease replaces at large currents monotonous growth which is seen at small currents. In the next section we show that different current dependences correspond to different detection scenarios.

### C. Current dependences of the jitter for deterministic and probabilistic scenarios

We invoke position dependence of the detection current across the nanowire [25] to evaluate current dependence of the intrinsic jitter in the deterministic regime. The authors of Ref. 25 showed that due to current crowding the detection current, $I_{det}$, depends on the hotspot position across the nanowire. The detection current reaches the maximum in the middle of the wire and drops to symmetric minima which are located near both edges of the wire. We denote corresponding values of the detection current as $I_{det}^{max}$ and $I_{det}^{min}$. The detection criterion is fulfilled when the velocity of superconducting electrons reaches the critical value. This occurrence initiate vortex crossing. Although at $I_B > I_{det}^{max}$ the whole wire responds deterministically, the size of the hot spot needed to initiate vortex crossing is different at different locations. It is larger in the middle of the wire and goes to minima close to wire edges. We note here that similar approach was used in Ref. 12 where the criterion for vortex crossing was suppression of the potential barrier. At $I_B < I_{det}^{min}$ detection criterion is not reached and the wire respond probabilistically. At intermediate currents, only that part of the wire where $I_{det} < I_B$ responds deterministically. Since the size of the hot-spot is largely undefined, we use the model of the diffusive normal spot [26] to connect the local detection current and the radius of the spot, $R$, as $(1 - I_{det}/I_{dep}) = 2R/w$, where $I_{dep}$ is the depairing current. The spot grows due to diffusion and reach the radius $R$ after a time $\tau_D = R^2/4D$ where $D$ is the diffusion coefficient. This time represents the time delay between photon absorption and opening the time window for vortex crossing. We further associate the standard deviation $\delta\tau$ in this time with the difference between diffusion times corresponding to the maximum and the minimum values of the detection current to obtain

$$\delta\tau = \eta^2 \frac{w^2}{16\,D} \left\{ \left(1 - I_{det}^{min}/I_{dep}\right)^2 - \left(1 - I_{det}^{max}/I_{dep}\right)^2 \right\}. \quad (6)$$

The coefficient $\eta$ accounts for uncertainty of the hot-spot size in the simplified hard-core model and will be used as fit parameter. We used the output of numerical calculations [25] to find the values of $I_{det}^{max}$ and $I_{det}^{min}$ at different mean values, which we associated with applied bias current, and computed the standard deviation. Fig 6 shows the best fit of the current dependence of $\sigma_{wire}$ at 800 nm. It was obtained with Eq. (6) for $I_C/I_{dep} = 0.88$ and $\eta = 2.2$. This value shows that the size of the hot spot, which is relevant for our model, is approximately twice as large as the size predicted by the hard core approximation. We have to note that Gaussian shape of PDFs which was found in the experiment can not be explained in the framework of our model. The most plausible reason is that the connection between local detection current and the size of the hot spot may also depend on the spot location. The approach of the potential barrier for vortex crossing also fails to reproduce Gaussian PDFs which we obtained in the experiment.

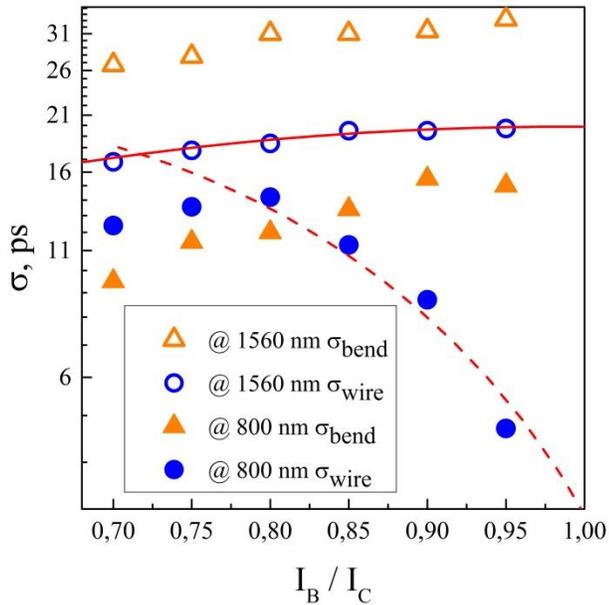

FIG. 6. (Color online) Contributions to the local jitter from bends (squares) and wires (circles) vs relative current. Open and closed symbols correspond to wavelengths 800 nm and 1560 nm, respectively. Lines show best fits to experimental data obtained with Eq. (7) (solid line) and Eq. (6) (dashed line).

If the deterministic detection criterion is not reached, the wire may respond probabilistically. In this case the vortex crossing occurs within the life time of the hot spot, $\tau_{HS}$. The delay between the spot appearance and the crossing start-time is described by exponential PDF with the mean rate $p = \tau_V^{-1}\exp[-U(I_B)/kT]$ where $U(I_B)$ is the current dependent instantaneous barrier for vortex crossing and $\tau_V \propto I_B^{-1}$ is the time it takes for the vortex to cross the wire in the absence of the barrier. Such statistics is typical for any system escaping from a metastable state over potential barrier [27, 17]. For NbN wires studied here, $\tau_V(I_C) = 12\ ps$. Because of the small size of the hot spot as compared to the wire width, change in the barrier $U$ during the time $\tau_{HS}$ is not uniform across the width. We simplified the approach by taking the expression for the current dependence of the barrier $U(I_B)$ with the uniform free energy (Eq. (13) from Ref. 17) and use the energy scale $\varepsilon_0$ as fitting parameter. In the framework of this approximation, the standard deviation in the delay time takes the form

$$\delta\tau = \int_0^{\tau_{HS}} (t - \langle\tau\rangle)^2\, p\, e^{-p\,t} dt, \qquad (7)$$

where the mean value $\langle\tau\rangle$ is defined with the same PDF within the life time of the hot spot. Best fit of the current dependence of $\sigma_{wire}$ at 800 nm with the Eq. (7) is shown in Fig. 7. It was obtained with $\tau_{HS} = 65$ ps and the value of $\varepsilon_0$ equal to 0.6% of its equilibrium value at the ambient temperature. The best fit value of the life time is more than three times larger than the life time 20 ps [22] concluded from correlation measurements. The difference between times translates into the 1.8 times difference in the size of the hot-spot. This is close to the 2.2 times difference in the hot-spot size required to get the best fit in the deterministic regime.

The fact that equations (6) and (7) fit our experimental data very accurately justifies our interpretation of two different contributions to the local jitter as contributions of deterministic and probabilistic events. This is especially sound for straight wires where the change from probabilistic to deterministic scenario with the increase in the current is clearly seen. Since for both wavelengths current dependences of $\sigma_{bends}$ are similar to the probabilistic current dependence of $\sigma_{wire}$ we conclude that bends detect photons probabilistically at any wavelengths. The reason is that the difference between $I_{det}^{max}$ and $I_{det}^{min}$ in bends is much larger than in wires [28] and true deterministic regime for bends can not be realized within available range of bias currents.

## V. CONCLUSION

Studying current dependences of timing jitter in photon counts delivered by meandering nanowires, we have found that not only the magnitude of the jitter but also its current dependences differ drastically at different photon energies. Moreover, we have found that photon count occurring in straight wires and bends demonstrate similar difference in current dependences of the jitter.

Analyzing statistics of appearance times of photon counts, we have shown that each count is a composite event including at least two elementary events described by different probability densities. We associated these events with the growth of the hot spot and random jumps of magnetic vortices across the wire. Depending on the current and photon energy, dispersion inherent to one of these two events dominates the jitter.

Finally, we have proposed simplified but analytical models which describe experimentally observed current dependences of timing jitter at different photon energies.

_______________________________________________